\documentstyle[preprint,epsfig,aps]{revtex} 
\def\ra{\rightarrow}
\def\be{\begin{equation}}
\def\ee{\end{equation}}
\def\bea{\begin{eqnarray}}
\def\eea{\end{eqnarray}}
\def\lln{\left<}
\def\rln{\right>}
\def\gam{\gamma}
\def\prl{Phys. Rev. Lett.~}
\def\pr{Phys. Rev.~}
\def\pl{Phys. Lett.~}
\def\np{Nucl. Phys.~}
\def\prl{Phys. Rev. Lett.~}
\def\tp{\vec v_1 \cdot (\vec v_2 \times \vec v_3)}
\begin{document}
\title{Triple product correlations in $B^0 \ra \bar \Lambda p 
\pi^-$}

\author{S. Arunagiri\footnote{arun@phys.nthu.edu.tw} and C. Q.
Geng\footnote{geng@phys.nthu.edu.tw}}

\address{Department of Physics, National Tsing Hua University,\\
Hsinchu, Taiwan 300, ROC}

\date{\today}

\maketitle
\begin{abstract}
Triple product correlations (TPC's) involving strange quark spin are 
elucidated in $b \ra u \bar u s$ process within the standard model. 
They arise when light quark masses are nonzero. As the momenta and spins of 
constituent quarks are related to that of the parent hadron, the quark 
masses are, however small, important and relevant in TPC studies. At this 
level the TPC's of interest are of the form 
$\vec s_s \cdot (\vec p_u \times \vec p_{\bar u})$ and 
$\vec s_s \cdot (\vec p_s \times \vec p_u)$. As an application, 
we look at $T$-odd violating effects in 
$B^0 \ra \bar \Lambda p \pi^-$ through the TPC
$\vec s_{\bar \Lambda} \cdot (\vec p_{\bar \Lambda} \times \vec p_p)$ 
for which triple product asymmetry is found to be 5.7--7.6\% in the 
vanshing limit of strong phase. 
\end{abstract}

\pacs{11.30.Er, 13.25.Hw}

\section{Introduction}
\label{sec1}
CP violating effects are sought after as to get the idea on the origin of
CP violation. In that pursuit, most of interests are now focussed on  
in $B$ decays which are expected to exhibit CP violation 'visibly'.  A 
component of CP violation in the CKM framework, namely 
$\sin 2\beta$, has already been measured by Belle and Babar collaborations 
at KEK and SLAC respectively \cite{belbab}. In these studies, 
both theory and experiment, the objective now turns out to be three 
folded: to test the CKM paradigm of CP violation, fix its limitations and 
to unfold the physics beyond it. 
  
Characteristic observables of CP violation are rate asymmetries 
and momentum correlations. The CP asymmetries, mixing induced 
and/or direct, arise if both the weak ($\phi$) and strong ($\delta$) 
phases are non-vanishing
\be
A_{CP}  \propto \sin \phi \sin \delta.
\label{cpa}
\ee
Whereas the correlations among spin and momenta of the intial and final 
state  particles constitute a measure of $T$-violating observables which 
implies CP violation by CPT theorem. The correlations known as 
triple product correlations (TPC's), of the $T$-odd form
$\vec v_1.(\vec v_2 \times \vec v_3)$, where $\vec v_i$'s are spin or 
momentum, are used to probe $T$-violation, for early works see 
\cite{okun,valen,geng} in $K$ as well as $B$ decays. Existence of a 
nonzero TPC is given by
\be
A_T =
{\Gamma (\tp > 0) - \Gamma (\tp < 0) 
\over {\Gamma (\tp > 0) + \Gamma (\tp < 0)}} 
\label{atp}
\ee
where $\Gamma$ is the decay rate of the process in question. 
In comparison with the 
conjucate process, TPC asymmetry (TPA), ${\cal A}_T$ is expressed 
as
\be
{\cal A}_T = {1 \over 2}(A_T-\bar A_T).
\label{tpa}
\ee
By expressing so, we reaffirm the TPC is indeed due to weak phase. 
Otherwise, the 
nonzero TPC in eq. (\ref{atp}) can occur due to only strong phase.
Then TPA turns out to be: 
\be
{\cal A}_T \propto \sin \phi \cos \delta.
\ee
This is in contrast with the CP asymmetry, eq. (\ref{cpa}).
TPA is $\it {protected}$ from strong interaction effects encoded in 
the phase, $\delta$. In the vanishing limit of the strong phase,  
the TPA is maximal, see \cite{london,london0}. We note that there is no 
contribution to ${\cal A}_T$ in eq. (\ref{tpa}) from final state 
interaction due to electromagnetic interaction. 

Consider the underlying quark process $b \ra 
u \bar u s$. The TPC's are among four momenta and four spins. Going on 
to hadronic level, the choice of correlation among them depends of upon 
the operators in the effctive Hamiltonian that decide the hadronic process. 
As regards inclusive decay, it is shown that the dominant 
operators being the tree and the penguins representing chromo electric and 
magnetic dipoles \cite{hou}, the non-negligible TPC's are 
$\vec p_{u{(\bar u)}} \cdot (\vec s_u \times \vec s_{\bar u})$ and 
$\vec s_b \cdot (\vec p_u \times \vec p_s)$. 
Example of hadronic process:  the former is applicable 
to $B \ra VV$, while the latter to $\Lambda_b \ra \Lambda \pi^+ \pi^-$. 
With the contribution of the operators mentioned, there is no TPC involving 
$s$-quark spin if strange quark mass being zero. It is, thus, argued in 
\cite{london} that an observation of $s$-quark spin heralds new physics. 
This statement is misleading and there are TPC's with $s$-quark spin if 
the strange quark mass in nonzero.  Does it matter? We address this question 
in this paper.

We know the operators that constitute the full effective Hamiltonian for 
the process $b \ra u \bar u s$ are many, see  \cite{buras,ali}. 
Of them, we consider a set of operators, which are different from the ones 
mentioned 
in the previous paragraph except the tree operators, that underly a 
hadronic process. These operators may be 
subdominant in comparison. Albeit, they completely determine a  
hadronic process. As we show below, there arise 
TPC's involving $s$-quark spin, $s_s$, within the standard model (SM): 
$\vec s_s \cdot (\vec p_u \times \vec p_{\bar u})$ and 
$\vec s_s \cdot (\vec p_u \times \vec p_s)$.
We hasten to note that the former occurs with nonzero $s$-quark mass 
and the other with that of $u$ quark mass. Thus, retaining light quark 
mass terms is to exhaust all the possible TPC's within SM.  
  
As an application, we consider the recently observed hadronic process
$B^0 \ra \bar \Lambda p \pi^-$ looking for TPC of the type 
$\vec s_{\bar \Lambda} \cdot (\vec p_{\bar \Lambda} \times \vec p_p)$. 
It is interesting on its own to note the branching ratio of three body 
baryonic decay in comparison with the same baryon pair in the two 
body final state as observed:
\bea
&&Br(B^0 \ra \bar \Lambda p\pi^-) = 
(3.97^{+1.00}_{-0.80}\pm0.56) \times 10^{-6}\hspace{0.6in} 
(BELLE\cite{belle})\label{exbr}\\ 
&&Br(B^+ \ra \bar \Lambda p) < 2.6 \times 10^{-6} \hspace{1.87in} 
(CLEO\cite{cleo})\\
&&Br(B^+ \ra \bar \Lambda p) < 2.2 \times 10^{-6} \hspace{1.72in} 
(BELLE\cite{belle0})
\eea
Such an enhancement of three body decay over the two body one is due 
to the reduced energy release in $B$ to $\pi$ transion by the 
fastly recoiling $\pi$ meson that favours the dibaryon production 
\cite{soni}. Theoretical estimations are made, in  consistent 
with the experimental observation, in a model dependent way \cite{chua}
\footnote{See \cite{group} for other interesting aspects of baryonic $B$ 
decays.}. We find the TPA in this process is at percent level in SM, 
following eq. (\ref{tpa}). This can be achieved 
at the $B$ factories in the near future with improved statistics of $B\bar 
B$ pair.

\section{Triple Product Correlations in \lowercase{$b \ra u \bar u s$}}
\label{sec2}
The effcetive Hamiltonian for $b \ra u \bar u s$ consists of many operator 
structures. Of them, the dominant operators are due to tree and 
chromo magnetic and electric dipoles (penguins). At the application 
level, choice over the quark operators depends upon the hadroic decay in 
question. The operators that enter 
the effective Hamiltonian for the hadronic processes we are interseted in 
are the tree level operators and the QCD penguin operators:
\bea 
Q_1 = c_1 (\bar u_\alpha \gam_\mu (1-\gam_5)b_\alpha)(\bar s_\beta 
\gam^\mu(1-\gam_5)u_\beta)\label{o1}\\ 
Q_4 = c_4 (\bar s_\alpha  \gam_\mu (1-\gam_5)  b_\alpha)(\bar 
u_\beta  \gam^\mu (1-\gam_5) u_\beta)\label{o4}\\ 
Q_6 = c_6 (\bar s_\alpha  \gam_\mu (1-\gam_5)  b_\beta)(\bar u_\beta 
 \gam^\mu (1+\gam_5) u_\alpha)\label{o6}
\eea
The QCD penguins $Q_{4,6}$ are subdominant with respect to those 
penguin-type dipole terms
\bea
&&Q_{g1} = c_{g1} (\bar s t_\alpha  \gam_\mu (1-\gam_5)  b)(\bar 
u t^\alpha \gam^\mu  u)\label{og1}\\
&&Q_{g2} = c_{g2} (\bar s t_\alpha  \sigma_{\mu \nu} q^\nu (1+\gam_5)  
b)(\bar u t^\alpha \gam^\mu  u_)\label{og2}
\eea
In all equations above $c_i$ are short distance Wilson coefficients, $Q_1$ 
corresponds to tree diagram. The coefficients of QCD penguin operators are 
less by an order of magnitude than the the coefficient of the dipole 
operator, see Ali {\it et al.} \cite{ali}.

On calculating the amplitude-squared of the process, one would find 
$T$-odd TPC's arise in the interference terms $Re(Q_1Q_4^\dagger)$ and 
$Re(Q_1Q_6^\dagger)$ but not in $Re(Q_4Q_6^\dagger)$. 
Since, in the interested hadronic process, the intial state being 
the meson, we sum over $b$-quark spin. Also, spins of $u$ and $\bar 
u$ quarks, which we are not interested in, are summed over. Their 
presence does not influence the determination of $s$-quark spin 
constituted TPC. As our interest lies over the qusetion of $s$-quark spin 
constituted TPC, we keep all quark masses nonzero to begin with. 
At this point, we disregard the sensitivity of the masses of the light quarks, 
especially $u$ quark mass. 

The $T$-odd terms with strange quark spin, $s_s$, are
\bea
Re(O_1O_4^\dagger)_{T-odd} &\propto& 
m_si\epsilon_{\mu\alpha\beta\gamma}p_b^\mu
p_u^\alpha p_{\bar u}^\beta s_s^\gamma\label{tpc1}\\
Re(O_1O_6^\dagger)_{T-odd} &\propto& 
m_{u}i\epsilon_{\mu\alpha\beta\gamma}p_b^\mu
p_u^\alpha p_s^\beta s_s^\gamma \label{tpc2}
\eea
We kept up only the leading terms in the respective interference terms, 
meaning that the terms proportional to $m_sm_u$, and like, are neglected.
Now we have TPC that involves $s$-quark spin within SM. At the rest frame 
of $b$ 
quark, we have TPC's of the form 
$\vec s_s \cdot (\vec p_u \times \vec p_{\bar u})$ and 
$\vec s_s \cdot (\vec p_u \times \vec p_s)$.  

The authors of Ref.\cite{london} considered the domiant operators, namely, 
$Q_1, Q_{g1}$ and $Q_{g2}$ in the vanshing limit of light quark ($u, d, 
s$) masses. As a result, there is no TPC with the spin of strange quark. 
Let us note that the spins and/or momenta of the constituent quarks would 
have, only upto a limited extent, bearing on that of the parent hadron. 
In the presence of light quark masses, there arise TPC's 
involving $s$-spin. For example, for $Re(Q_1Q_{g1}^\dagger)$, we have two 
TPC's same as in eqs. (\ref{tpc1}) and (\ref{tpc2}).

On the question over choosing to retain the light quark masses. In the 
vanishing limit of light quark masses, there arises no TPC with strange 
quark spin. Any follow up statement is, then, unfounded. Keeping the masses 
of light quarks is to serve the purpose if particular TPC at quark level 
exists or not in SM. This is the objective of working at quark level. The 
light quark masses do not, anyway, influence the magnitude of TPC at 
hadron level. That is the magnitute of quark level TPC's (qTPC's) are not 
expected to be carried over to hadronic level. Some of qTPC's would 
survive up to hadron level while some would disppear during hadronisation.

Both the TPC's 
$\vec s_s \cdot (\vec p_u \times \vec p_{\bar u})$ and 
$\vec s_s \cdot (\vec p_u \times \vec p_s)$
can be identified with  
$\vec s_{\bar \Lambda} \cdot (\vec p_{\bar \Lambda} \times \vec 
p_{p,\pi})$
in $\bar {B^0} \ra \Lambda \bar p \pi^+$
and $\Lambda_b \ra \Lambda \pi^+ \pi^-$, 
since quark model that relates hadron to its constituent quarks does 
not distinguish $p_u$ and $p_s$ with regard to $\Lambda$. There also 
appear, as pointed out in Conclusion, qTPC's with $b$-quark spin that 
corresponds to the spin of $\Lambda_b$, if the spin of $b$-quark is 
considred. In this note, we consider the former decay mode only.   

\section{$B^0 \ra \bar \Lambda p \pi^-$}

Applying equations (\ref{o1}-\ref{o6})
and by factorisation 
approximation, the invariant amplitude for $B^0(p) \ra \bar \Lambda 
(p_{\bar \Lambda},s_{\bar \Lambda}) 
p(p_p) \pi^-(p_\pi)$, where $p$'s are the momenta and $s$ the spin, 
consists of \cite{chua}: 
\bea
M_1 &=& {G_f \over \sqrt{2}} V_{ub} V_{us}^* c_1 
\lln \pi^+ | \bar u \gamma^\mu(1-\gamma_5)b|\bar {B^0}\rln
\lln (\bar \Lambda p|\bar s \gamma_\mu(1-\gamma_5)u|0\rln e^{i\delta_1}  
\label{m1}\\
M_4 &=& -{G_f \over \sqrt{2}} V_{tb} V_{ts}^* c_4 
\lln \pi^+ | \bar u \gamma^\mu(1-\gamma_5)b|\bar {B^0}\rln
\lln (\bar \Lambda p|\bar s \gamma_\mu(1-\gamma_5)u|0\rln e^{i\delta_4}
\label{m4}\\  
M_6 &=& \sqrt{2}G_f  V_{tb} V_{ts}^* c_6 
\lln \pi^+ | \bar u \gamma^\mu(1-\gamma_5)b|\bar {B^0}\rln {(p_\Lambda 
+\bar p)_\mu \over {m_b - m_u}}
\lln (\bar \Lambda p|\bar s (1+\gamma_5)u|0\rln 
e^{i\delta_6}\label{m6}
\eea  
That the $B^0 \ra \pi^-$ transition accompanied by the current 
produced $\bar \Lambda p$. One that $B^0 \ra \bar \Lambda 
p$ with current produced $\pi^-$ does not contribute. That is the 
absence of operator $O_2$. This is 
analogous to $\bar B^0 \ra K^{*-}\pi^+$ where $\bar {B^0} \ra K^{*-}$ 
accompanied by $\pi^+$ does not occur. 

With the amplitude-squared being
\be
|M|^2 = |M_1|^2 + |M_4|^2 + |M_6|^2 + 2 Re (M_1 M_4^\dagger) + 
2 Re (M_1 M_6^\dagger) + 2 Re (M_4 M_6^\dagger),
\ee
$Re(M_1 M_4^\dagger)$ and $Re(M_1 M_6^\dagger)$ have the $T$-odd terms. 
Let us now note the absence of TPC in $Re(O_4O_6^\dagger)$ as well as 
$Re(M_4M_6^\dagger)$ which 
demonstrates a one-to-one correspondence of a qTPC at 
quark level to the 
hadron level. Therefore, the absence of a hadronic level TPC at quark 
level is a signal of possible new physics. 

Then, the width associated with the TPC 
$\hat {s}_{\bar \Lambda} \cdot (\hat {p}_{\bar \Lambda} \times 
\hat {p}_p)$ is
\be
\Gamma_{TPC} = X Im[V_{ub}^*V_{us}V_{tb}V_{ts}^*e^{i \delta}]
 \label{gxy}
\ee
where
\bea
X &=& 4 G_f^2|F_1^{B \ra \pi}(q^2)|^2 m_B^2 m_\Lambda m_p\nonumber\\
    && \times \left\{-c_1c_4g_Ah_A+c_1c_6[(F_1+F_2)f_S+g_A g_P] 
{m_\Lambda+m_p \over {m_b-m_u}}\right\}.
\eea

In $X$, the brayonic form factors $g_A = -1.45913, h_A = 0.71037, 
g_P = - 0.81561, f_S = g_P, F_1 +F_2 = -0.27373$ (see 
Ref. \cite{chua} and reference there in for details), $F_1^{B \ra 
\pi}(q^2) = 
F_1(0)/[(1-q^2/M_V^2)(1-\sigma_1q^2/M_V^2)]$ with $F_1(0) = 0.29, 
\sigma_1 = 0.48, M_V = 5.32 GeV$ and the Wison coefficients $c_1 = 1.117, 
c_4 = -0.044, c_6 = -0.056$. In calculating all the form factors, $q^2 = 
(m_\Lambda+m_p)^2$ is used.

The phase factor is, in terms of Wolfenstein parameters,
\be
Im[V_{ub}^*V_{us}V_{tb}V_{ts}^*e^{i \delta}]
= - A^2\lambda^6[\eta \cos \delta + \rho \sin \delta].
\ee
Alongwith the conjucate one, we would get the $sine$ term cancelled out 
making TPA proportional to $cosine$ term. Vanishing limit of strong phase
gives rise to maximum CP violation. 

On doing phase space integration using RAMBO\footnote{We thank Prof 
David London for providing this program.}, for vanishing 
strong phase and $\eta$ being 0.3 to 0.4 \cite{pdg}, the TPA is, as given 
by eq. (\ref{tpa}),  
\be
{\cal A}_T \equiv 5.7\--7.6\% \label{ntpa}
\ee
In obtaining the numerical value of ${\cal A}_T$, we 
have used the central value of the observed decay rate, vide 
eq. (\ref{exbr}) that has been consistently accounted for by the 
invariant amplitude in eqs. (\ref{m1}-\ref{m6}) which is model 
dependent that would have some influence on TPA in eq. (\ref{ntpa}).
The TPA obtained above is maximal. In view of the wide belief that the 
strong phase is quite small, we can expect the TPA still being at percent 
level with actual strong phase. {\it Any observation of the TPA beyond 
percent level as obtained here, vide eq. (\ref{ntpa}), would be a signal of 
new physics}.

\section{Conclusion}       
As has been emphasised, $T$-odd violation is a window in 
looking for CP violation in $B$ decays. The $T$-violating effects are 
expressed through TPA in analogous with CP asymmetry. The disticntion is 
that the TPA gets maximal in the vanishing limit of strong phase. 

In order to see the implication of experimental results at quark level, we 
have to know the connection between the TPC variables, namely, spin and 
momentum, of hadrons and that of constituent quarks. At least for baryon, 
we believe there is an established connection. For example, spin of 
$\Lambda_b$ and $\Lambda$ with that of $b$ and $s$ quarks respectively.     
In the constiuent quark picture, we cannot particularly link the momentum 
of a hadron to any of the constituent quarks which are all supposed to 
carry equally shared momentum of the hadron. 

Thus, we looked at qTPC's in $b \ra u \bar u s$. We found TPC 
involving strange quark spin of the form 
$\vec s_s \cdot (\vec p_u \times \vec p_{\bar u})$ and 
$\vec s_s \cdot (\vec p_u \times \vec p_s)$. These TPC's exist within 
SM only when light quark masses are non-vanishing. In application to 
hadronic process, $s_s$ is related to $s_\Lambda$ and the momenta to any 
light hadron such as $\Lambda, p, \pi$ etc. as applicable. 

We have also observed there is a  one-to-one correspondence between quark 
level and hadronic level TPC's. In order to identify if a particular 
hadronic TPC has its equivalant one at quark level, the (light) quark 
masses have necessarily to be retained. Then only, physics beyond standard 
model in terms of TPC can be enunciated, noting that quark level TPC might 
be absent at hadronic level and the converse does not hold as far as weak 
interaction is concerned. In other words, in SM, a hadronic level TPC 
should have its conterpart at quark level and any 
absence would be a smoking signal of possible new physics. In order to 
ascertain the absence or presence of qTPC, it is necessary to keep the 
mass terms irrespective their magnitude.

In view of this, it is now clarified that {\it strange quark spin 
constitutes TPC within the standard model}. We have shown there are two 
TPC due to $s$-spin. It is irrelevant how suppressed at quark level the 
corresponding term is on account of the presence of light quark mass. 
What is significant is the existence of TP variable at quark level. 

As an application, $B^0 \ra \bar \Lambda p \pi^-$ is considered. 
The $T$-violating effects are looked at through the TPC
$\vec s_{\bar \Lambda} \cdot (\vec p_{\bar \Lambda} \times \vec p_p)$. It is 
connected to  $\vec s_s \cdot (\vec p_{s,\bar u} \times \vec p_{u})$ at 
quark level. It is found that the TPA is about 5.7--7.6\% in SM in the 
vanishing limit of strong phase. Hadronic level TPC has the corresponding 
qTPC. Thus, if TPA to be abserved exceeds, it then 
signals new physics.

Further application can be made for the process\footnote{See \cite{gengg} 
for TPC in $\Lambda_b \ra \Lambda l^+l^-$.} $\Lambda_b \ra \Lambda 
\pi^+ \pi^-$ with the TPC being $\vec s_\Lambda \cdot (\vec p_\Lambda \times 
\vec p_\pi)$ which has the counter part at quark level in both of 
eqs. (\ref{tpc1}) and (\ref{tpc2}). With $b$ quark polarisation 
corresponding to that $\Lambda_b$, there exist qTPC's of the form 
$\vec p_{u(\bar u)} \cdot (\vec s_b \times \vec s_s)$ and  
$\vec p_s \cdot (\vec s_b \times \vec s_s)$. The corresponding TPC at 
hadronic level is 
$\vec p_{\Lambda,\pi} \cdot (\vec s_{\Lambda_b} \times \vec s_{\Lambda})$.
It is interesting to see the TPA in this process.

Finally, we note that in order to observe the TPA being at 5.7--7.6\%, 
we need to have about (4.3--7.7)$\times 10^{7}$ $B\bar B$ 
pair at $1\sigma$ level. This is within the reach of the present day $B$ 
factories at KEK and SLAC and others that would come up.  

\section{acknowledgements}
This work is financially supported by the National Science Council of 
Republic of China under the contract number NSC-91-2112-M-007-043.

\end{document}